# New contexts, old heuristics: How young people in India and the US trust online content in the age of generative AI

Rachel Xu

Google Jigsaw, rachelxu@google.com

Nhu Le

Gemic, nhu.le@gemic.com

Rebekah Park

Gemic, rebekah.park@gemic.com

Rebekah Park

Gemic, rebekah.park@gemic.com

Laura Murray

Gemic, laura.murray@gemic.com

Vishnupriya Das

vishnupriya.das@gmail.com

Devika Kumar

Gemic, devika.kumar@gemic.com

Beth Goldberg

Google Jigsaw, bethgoldberg@google.com

We conducted in-person ethnography in India and the US to investigate how young people (18-24) trusted online content, just as generative AI (genAI) became mainstream. We found that when online, how participants determined what content to trust was shaped by emotional states, which we term "information modes." Our participants reflexively shifted between modes to maintain "emotional equilibrium," and eschewed engaging literacy skills in the more passive modes in which they spent the most time. We found participants imported trust heuristics from established online contexts into emerging ones (i.e., genAI). This led them to use ill-fitting trust heuristics, and exposed them to the risk of trusting false and misleading information. While many had reservations about AI, prioritizing efficiency, they used genAI and habitual heuristics to quickly achieve goals at the expense of accuracy. We conclude that literacy interventions designed to match users' distinct information modes will be most effective.

**ACM Reference Format:** Reference to come

## 1 INTRODUCTION

The introduction of genAI has challenged the ways people have learned and discerned truth in the past two decades of online information sharing. How people have adapted to this reality; what strategies they employ to evaluate online information; and how these new strategies interact with established and well-researched heuristics—such as source reputation, content aesthetics, and comments as crowdsourced wisdom—were animating questions for this research.

Established literature on the subject of online trust, heuristics, and social harm are dominated by the fields of psychology, marketing, and communications [Kahneman 2011, Metzger 2010, Petty 1986, Sundar 2008]. Scholars in these fields contest how trust works in important and nuanced ways, and yet, there is often a shared presumption that the individual in isolation is the most important unit of study [Keib 2017, Lichtle 2015, Uddenberg 2023, Shen 2018]; human cognitive processing works in ways that are timeless and universal [Herfeld 2021]; and misinformation is a clear matter of true and false statements [Hwang 2021, Pennycook 2019]. Against this backdrop, our work investigates real-time practices of trusting as they (1) unfold in relation with other people, groups, and identities; (2) change over time with the introduction of new technologies; and (3) adapt to evolving social truths about the nature of authority, expertise, and credibility. That is, using ethnographic methods, we foreground the social, dynamic, and adaptive nature of trust heuristics in order to shed new light on what is at stake in the changing ways that young people trust online content today [Compton 2021, Lewandowsky 2017, Qian 2022]. Specifically, we ask:

(RQ1) What trust heuristics did young people use when engaging online information across search engines, social media, direct messaging and interactions with generative AI?

(RQ2) How did young people perceive AI-generated content and assess its trustworthiness?

This paper offers an ethnographic perspective on when trust matters in digital settings, what it looks like in practice, and the potential harms that are created when heuristics of trust that evolved in one context (i.e. social media posts) are carried into another (i.e. genAI). Our research produced five findings:

1. When online, participants were engaged in different mindsets and emotional states, which we term "information modes," that they rapidly and unconsciously switched between to maintain an emotional equilibrium. They avoided what felt emotionally taxing and sought out what felt energizing and soothing.
2. The trustworthiness of the content only mattered in some modes. In the most popular modes where they spent most of their time, they overrode their literacy skills.
3. Participants readily adopted genAI text assistants to get ahead in their future careers and saw genAI as just another tool to master. This desire was driven by anxiety about the future in the US, and about excitement about the future in India. This mental model and bias towards action exposed them to potential harms such as misinformation online.
4. Participants applied outdated trust heuristics from other online interactions (such as search engines and social media) to their online interactions with genAI, which led them to overestimate genAI and themselves.
5. Participants viewed genAI as a safe space without social risk or judgment. Many participants expressed fear of social sanction or "being canceled," being embarrassed for not knowing the answer, or taxing their social connections for speaking. GenAI served as a safe alternative to ask questions and test ideas without consequences.

Our findings suggest that young people are maneuvering between different modes when online – and that many of these modes involve a turn away from traditional media literacy skills. Rather than attempt to stem this tide, the HCI community can support young adult explorations of genAI by (1) designing products and literacy interventions that meet young people in the modes they are in, and (2) building upon their existing fact-checking instincts and adaptations.

## 2 LITERATURE REVIEW

### 2.1 Evolution of truth and online trust practices

Scholars in both the US and India have studied the ways people use technology to construct and broker truth online, finding that technological change brings about shifts in trust paradigms [Ali 2021; Guess 2020; Grant 2011; Hanckel 2021]. At the same time, studies on media repertoires, or the range of media sources available to people, have highlighted the proliferation and increased importance of digital sources, especially user-generated content [Hasebrink 2012]. As a result, we've shifted away from trustworthiness being based solely on expertise to instead being based often on perceived relatability [Xu 2024] This shift speaks to the current moment being what Jayson Harsin [2018] described as a "post-truth regime:" a societal state in which information no longer needs to be based on institutional approval, academic pedigree, and scientific rationality—a conception of truth that in the West dates back to the Enlightenment—to be considered trustworthy.

This socio-technological context casts doubt on "information literacy" as a useful framework for conceptualizing online information seeking behaviors, or how people engage with online information. In their critique of the information literacy frameworks, metaliteracy experts Thomas P. Mackey and Trudi E. Jacobson point out that such approaches assume a discrete, linear model for information-vetting, wherein a single user exercises core competencies such as triangulation or fact-checking to evaluate information to determine the "truth" [2011\. In fact, real-world online information-seeking behaviors are often non-linear, non-discrete, and socially-motivated [Hassoun 2023].

In a post-truth era, it is thus necessary to approach information-seeking not as an individualized and rational process, but instead as a "social and connective act, performed in relation to collective norms and group identities" [Swart 2023: 520]. This suggests a need for a new way to understand how people engage with information. Our research proposes a new framework for information seeking: "information modes."

In putting forth this new model, we aim to make three contributions to the current literature:

First, "information modes" extends and builds upon the paradigm of "information sensibility" previously proposed by anthropologist Amelia Hassoun and colleagues as an alternative to information literacy. Hassoun and colleagues defined "information sensibility as information-seeking motivated by the "socially-informed awareness of the value of information." Our work furthers this definition by detailing the varied social motivations and contexts that meaningfully alter the value of information and therefore a user's information-seeking behaviors. These meaningful sets of motivations, contexts, and behaviors are what we call "information modes."

Second, we sought to identify new "trust heuristics," and the online social contexts (or information modes) under which they are deployed. Heuristics are, broadly, cognitive shortcuts that increase efficiency during rational reasoning [Kahneman 2011; Gasser 2012; Metzger 2013]. We define trust heuristics specifically here as conscious and unconscious time-saving strategies used by young users in each information mode to verify the credibility of content.

At study outset, our team prioritized three prominent trust heuristics that are examined at length within existing literature: (1) crowdsourcing credibility, or the probabilistic deliberation of content being true based on peer accounts or crowd wisdom [Hassoun 2023; Pfeuffer 2022]; (2) source credibility, or when the reputation of a source confers a sense of trustworthiness to the content itself [Fedeli 2018; Metzger 2003; Lim 2022; Xu 2024]; and (3) like- minded and like-bodied, or the bias toward content from a creator mirroring similar looks, values, and identities to them [Lee 2022; Karizat 2021; Xu 2024].

It is important to note that in our study and discussion of trust heuristics, we do not hold the position that heuristics involve cognitive trade-offs, which the prevailing literature has often over-emphasized [Petty 1986; Peng 2023; Jenkins 2020]. This view often moralizes the use of heuristics as an undesirable way of making judgments and simultaneously undervalues how heuristics frequently evolve in changing informational contexts. At this moment, when the pace of content production has rapidly increased and truth is regularly contested, heuristics help young people tackle new challenges that these socio-technological trends have introduced: namely, information overload and the social pressure to take a stance on every potential "truth" [Zhou 2022; Melumad 2019; Chen 2023]. From this view, heuristics are not inherently good or bad but are socially-situated practices for sense-making.

Third, we sought to understand how usage of genAI, as a new tool, interacts with existing and emerging info-seeking behaviors and trust paradigms. We use genAI here to refer to deep learning models that learn patterns and structure of input data to generate new synthetic data that resembles real-world text, images, audio, or video [Dwivedi 2023; Baidoo-Anu 2023; Morris 2023]. This is an evolution of narrow AI, which is AI that has been trained to complete specific tasks, and is also distinct from artificial generalized intelligence (AGI), which is a theoretical artificial intelligence with a human level of cognitive ability [Morris 2023].

Many discussions of trustworthiness of genAI tend to frame the topic as either an institutional matter, or as a capabilities and literacy matter. The former approach details technological, social, and regulatory challenges and hypothesizes or tests structural solutions to increase trust in genAI [Lenat 2023; Dunn 2023; Li 2023; Krishnaram 2023; Baldassarre 2023]. The latter approach focuses on how individuals understand what is and is not trustworthy, and what user-level interventions or technological affordances can boost their ability to understand and assess genAI trustworthiness [Xu 2023; Ali 2021].

The former approach often leaves individuals with too little agency, while the latter leaves them without context. And both approaches view trust in genAI tools and genAI content as a challenge that is unique and discrete from individuals' trust attitudes and practices with respect to the internet writ large. By divorcing social actors from the ability to change, influence, and shape cultural evolution, or conversely, removing them from the social, we lose the ability to understand why and how certain emerging and even marginal practices become social systems. In contrast, our approach, based on the principles of ethnography, empirically contextualizes trust in genAI in terms of existing online trust practices as well as broader social pressures and cultural beliefs.

The result is that we can better anticipate the future of trust and online info-seeking behaviors and its interactions with genAI. This will enable the HCI community to build better digital interventions to help young people navigate online information, especially on high-stakes topics such as election results.

## 2.2 Misinformation from Global South to North

Our cross-cultural research resisted positioning a Western context as the default in which to compare non-Western society and behavior. Instead we used an "India-first" framing – an approach designed to challenge the "west to rest" approach that dominates scholarship on tech-mediated information seeking behaviors [Athique 2020; Couldry 2021; Nithya 2021]. Doing so enabled our research team to develop analytic categories and hypotheses that were rooted in empirically observed events, utterances, and beliefs.

For example, we argue that "timepass" was a mode of navigating digital ecosystems that dominates how young people encounter and trust online information today. "Timepass" is an Indian idiom, and despite being a concept that is etic to American behavioral settings, it galvanized our understanding of how young people globally are finding, vetting, and trusting online information – or not.

More broadly, generating theory "from South to North" was an intellectual framing that helped surface signals from the non-West that "provincialized" taken-for-granted concepts in more general literature about heuristics [Chakrabarty 2000]. For example, Indian scholars have found that misinformation exists not because its creators or consumers lack literacy, but rather, because misinformation can create moments of consolidation of participatory cultures [Banaji 2021].

Thinking with India-first was therefore not an all-encompassing paradigm for our work, but rather, an important tool derived from anthropological theory to produce cross-cultural analysis that was empirical, novel, and reflective of the complex pathways that social evolution takes in a globalized present.

## 3 METHODOLOGY

Studying heuristics presents a methodological challenge because people gravitate toward, respond to, absorb, use, and pass on information in ways that are hard to consciously reflect on and articulate after-the-fact. To address this challenge, we drew from the "Manchester School," which emphasizes capturing particular instances over generalities to understand social phenomena [Evens 2022; Turner 1957]. Working from real-life settings, we honed in on precise empirical details of users' trust behaviors and used an iterative analytic approach to map how singular events can crystallize and illuminate broad socio-cultural realities. This unique approach helps capture information practices *in context*, leading us to findings that build upon those previous scholars of trust have come via deductive and experimental approaches [Hargittai 2010; Kaur 2022; Yu 2018].

### 3.1 Recruitment and Participant Profile

This study included 52 participants aged 18-24 years old, born between 1998 and 2004. We use the term "young people" to refer to study participants, but want to acknowledge that the original framing of our research was around "Gen Z." We recruited for Gen Z because of their reputation as "digital natives" [Bennett 2008, Vardeman 2022]. We wanted to study how those who think the assimilation of new technologies into daily life is unremarkable have responded to genAI.[1]

We had two field sites: Bangalore (*n* = 26) and New York metro area (*n* = 26*)*. These two cities were selected as similar Tier 1 urban areas with a reputation for a youthful population and technological sophistication, due to our interest in understanding emerging genAI practices amongst young people and its implications for trust.

Across both field sites, we included participants with varying levels of genAI literacy based on self-reported familiarity with and usage of genAI tools: 9 highly familiar, 42 familiar, and 3 unfamiliar.[2] We categorized participants based on their self-reported level of AI-literacy during the in-person interview, the second phase of the study. Since adoption of AI tools was very rapid and unpredictable during data collection, we did not conduct benchmark testing for AI literacy, and did not aim for a specific distribution of AI literacy in our sample to mirror the general population. Instead, we sought diverse representation for self-reported AI literacy.

---

[1] In writing this paper, however, we have opted to not use the term Gen Z because we did not study other generations in similar life stages. We removed generational terminology because our findings do not discuss distinct generational differences. This aligns with the approach demographic research institutions such as the Pew Research Institute take [Parker 2023].
[2] We defined each tier of AI literacy as follows: AI-unfamiliar participants did not know and had never used any generative AI tool; AI-familiar participants had previously used at least one generative AI tool, typically a text-based tool like ChatGPT; highly familiar participants were familiar with or had used multiple non-text based generative AI tools, such as DALL-E, in addition to regularly using a text-based tool.

We also recruited participants from locations with varied community density and tracked education and income levels; all were students or early-career adults. The recruitment and selection for cultural identity varied by cultural context. In the US, we selected for diversity in terms of race and gender. In India, we included a diversity of ethno-religious groups and participants who identified as either female or male (see Appendix Table 1). We used an online study recruitment platform, snowball recruitment, and posting invitations online.

### 3.2 Research Phases and Process

Data collection took place from May 2023 to July 2023 in three phases. This three-phase design enabled iterative rounds of data collection and analysis, which improved data quality because researchers were able to conduct interviews in Phase 2 with knowledge about participant biographies and beliefs gained from Phase 1. A phased approach also provided participants with the opportunity to validate or amend initial findings and provide feedback on interventions developed in Phase 3.

Our approach to data collection was thus designed for more longitudinal and reciprocal engagements between researcher and participant – aligning with ethnographic best practice by making research findings more robust and alleviating ethical concerns about research being alienating and incomprehensible to its subjects [Birt 2016] – while also working within the more condensed time frame required for studying an emergent and rapidly changing topic, like generative AI.

*Phase 1 - Journey mapping, diary study, and app history*

We completed a multidisciplinary literature review to develop effective recruitment tools and screening questions. We asked recruited participants to catalog: (1) what social media platforms and online spaces they visited that day; (2) what content they saw that felt "trustworthy" and "untrustworthy" and why. Participants uploaded screenshots and links, describing their perceptions of the content's credibility. We analyzed preliminary patterns and themes in short-answer responses from 130 people.

*Phase 2 - In-person ethnography*

We visited 52 participants' lived spaces to observe how their offline context influenced online practices. This enabled us to observe participants' unspoken trust attitudes and practices, which are often instantaneous and instinctive. We audio-recorded and transcribed all interview sessions and video-recorded specific segments. For each interview, we composed field notes and conducted thematic analysis of the interview data through coding [Saldana 2021].

*Phase 3 - Coding data and remote follow-up interviews*

We open coded our data from Phase 1 and Phase 2 [Holton 2007; Williams and Moser 2019]. We used preliminary findings from this first round of coding to develop ideas for potential information literacy interventions.[3] We conducted 60-minute follow-up interviews remotely to present these intervention ideas and collect additional data. We also discussed initial findings from our coded data with our participants. The research closed with collaborative analysis sessions to examine codes within and across thematic categories, developing new insights.

---

[3] The research team designed these interventions to address unmet needs in vetting practices or to help shed light into contradictions in the data. For example, we showed a mock screenshot of a fact-checking note on different platforms, which we postulated could be written by another user, a professional fact-checker, or generative AI. Participants explained why and how their reactions to the intervention differed by platform and source of the fact-checking, which enabled us to better understand source and platform as credibility cues.

### 3.3 Ethical Considerations

The study plan was reviewed by experts in human subjects research, legal, security, and privacy. We consulted academics and experts who specialize on topics of trust, genAI, and youth. All participants provided both written and verbal informed consent prior to commencing the study, and were reminded they could withdraw from the study at any time without consequences.

We maintained strict data privacy control for all participants. We assigned participants a pseudonym for all data collections. We use different pseudonyms in this paper from those used during collection and analysis. Participants were instructed to withhold personally identifying information (PII) during the study. Diary responses from Phase 1 and interview transcripts from Phase 2 and 3 were scrubbed of PII. Raw research data were deleted within 30 days of the final reporting.

### 3.4 Study Limitations

Our sample size was not statistically representative of the 18-24 year old age cohort. Our sample was limited to two geographic locations accessible to urban areas, which may not represent broader Indian and US youth experiences with genAI. For example, Bangalore is recognized for its technological sophistication, which means our sample may skew towards young people of higher technological proficiency, income, and education. As a result, the integration of genAI practices in everyday engagement with online information may not be as prevalent or pronounced in the general population as in our sample.

Self-reported data, as captured in all phases of the study, but particularly the digital diary in Phase 1, is known to have key limitations such as recall, observer, and social desirability biases. To mitigate these biases, we conducted multiple phases and incorporated multiple data collection methods, which enabled us to cross-reference data between each phase and ask probing questions if contradictions or ambiguities occurred. By capturing real-time reactions to online content, including body language, we minimized recall bias and social desirability bias.

GenAI technologies were evolving throughout the study, and users were still forming their expectations and mental models for AI products. Our findings on these should thus be treated as a snapshot in time rather than the permanent viewpoints of these users.

Finally, the authors and researchers do not share the same generational experiences and cultural backgrounds as participants. None of the authors and researchers are 24 years old or younger. In India, the field research was conducted by one Indian researcher and one Indian-American researcher, both of whom live in the US. These factors may have constrained the research team's ability to capture and understand the social context of participant answers. We mitigated this through regular research team discussions to actively reflect on the impact of our positionality throughout the research.

## 4 FINDINGS

This research was conducted at a critical juncture for young people's information journeys—at a time when genAI was first introduced to the market and young people were adapting to this disruption and incorporating it into their lives. In order to understand how genAI shapes young people's trust behaviors and heuristics, we first present findings about how their information seeking behaviors have evolved agnostic of any new tools (Section 4.1). Then, we consider the implications of introducing genAI alongside these new behaviors (Section 4.2).

### 4.1 Information modes as a new model for info-seeking

We found that participants sought and avoided content in terms of two key variables: content "weight" and content "sociality." Content weight refers to whether content felt emotionally "light" or "heavy." Content that had low-mental tax and was entertaining or soothing was considered light. Content that had high-mental tax and was boring or triggering was considered heavy. Content sociality refers to whether content impacts just the participant (self) or involves others (social). The former felt "obligation-free," while the latter conferred an obligation to act after they consumed it. For example, participants perceived news about war and politics as carrying social consequences and compelling them to act upon learned information.

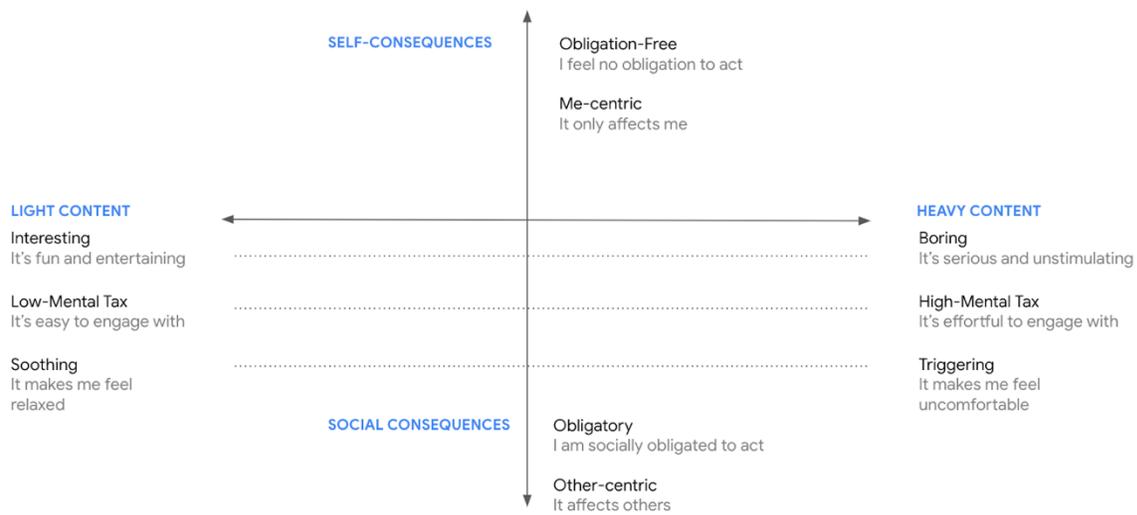

Fig 1. Participants perceived and reacted to the value of content along two key spectra

Based on this schema, we developed the concept of "information modes." "Information modes" are the emotional mindsets that people were in when engaging with online content. We emphasize emotion because participants did not consciously and rationally determine where content fell along the spectra, they reacted instinctively.

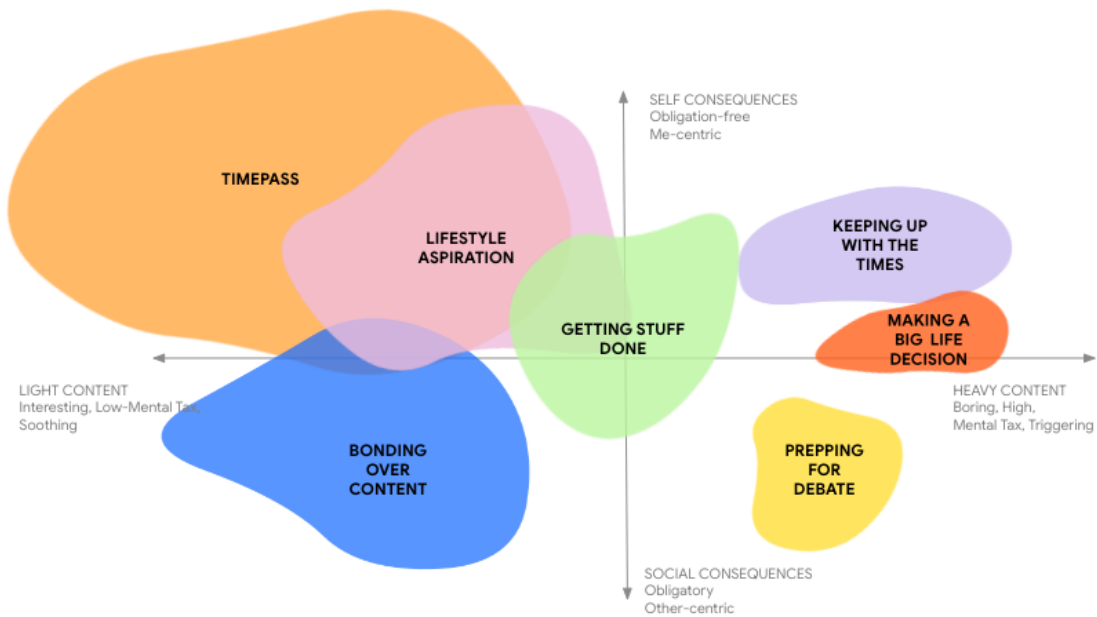

Fig 2. Map of info modes, sized to estimate proportion of time spent online, along key spectra

Kathy's (21, USA) description of how she consumed content was typical of our participants:

"[My feed] sends me videos like today's news or what's going on in pop culture or skincare or clothing … I don't personally seek out news … When you are constantly exposed to [news like school shootings], you feel more sad. You want to watch something that brings you joy or brings you entertainment … [Though] obviously you should be aware of what's going on in the world."

Kathy's quote highlights the way participants' intuitive attraction or repulsion shaped what other content they sought out or were fed by their feeds. It also highlights the different – sometimes conflicting – objectives that participants had for consuming content. These objectives made up the tapestry of their emotional mindset, and consequently, their information mode.

Unconsciously engaged information modes, in turn, determined trust behaviors. We observed participants moving between different information modes and trust contexts in quick, unplanned ways. They "wandered" in and out of different modes – and each mode shaped how and when they deployed different heuristics to trust online information.

We specifically identified seven key information modes participants inhabited, distinguished by their motivations for spending time in each mode and trust heuristics deployed (see Appendix B for full summary).[4] For example, in Timepass, where young people spend the most time, participants were looking to assuage boredom while maintaining their emotional equilibrium. Trust heuristics were irrelevant because the content was consumed for entertainment or soothing purposes, and checking veracity claims was seen as unimportant.

---

[4] We estimated this time based on screenshots participants shared of time spent on different apps, observation of their phone habits, written exercises, and their retrospective narrative during the interviews.

"If it's just for humor, the fun entertainment industry and all, you can contradict yourself." (Timepass Mode) – Gaurav (23, IND)

In Lifestyle Aspiration, participants sought information that would help them improve themselves and reach towards their ideals. As a result, they applied the "like-minded, like-bodied" trust heuristic, which is to say, they valued content stemming from personal experience, created by people who mirrored their identities and values because it felt most "applicable."

"[Andrew Tate] might be wrong, he might be right, but the way he is influencing young men in society is quite motivating actually. [Because of him] I go to the gym more, work hard more … I want to get that rich." (Lifestyle Aspiration) – Mehul (21, IND)

In Bonding Over Content, participants consumed and shared content with online communities to unlock a sense of belonging. They utilized the source credibility heuristic, leveraging peer ties to filter content. By opting into communities with content norms and values that they assumed were shared across all members, they conferred trust to all content shared within these communities.

"There's a sense of community in [sharing info]. With fandom, it's not harmful to hype yourself up over something that doesn't end up being true." (Bonding Over Content) – Avery (20, USA)

In Getting Stuff Done mode, participants were focused on accomplishing personal tasks efficiently, and were therefore trying to leverage online knowledge to work faster. The veracity of information was often considered less important than accomplishing their task. Information vetting was disregarded because it slowed them down.

"As long as I understand and I know what I'm talking about, at the end of the day it doesn't matter how I got there, especially if it's just like, stupid assignments." (Getting Stuff Done) – Irene (18, USA)

In Keeping Up with the Times, participants wanted to stay up-to-date on important topics. They applied heuristics that decreased the amount of time and energy they had to spend in these modes. For example, some participants described reading headlines only, but strictly from previously-vetted content creators or outlets.

"I will wait for a trusted source to post [about a news event] … Each news source might have a different take, like, it's complicated. So I just stick to the ones I trust." (Keeping Up with the Times) – Paresh (21, IND)

In Prepping for Debate, they wanted to learn about topics in-depth to feel safe from potential peer judgment in discussions. This was the only mode in which participants applied traditional information literacy skills despite it being time-consuming and emotionally-draining, because they needed to be able to state facts and defend positions. As a result, they minimized the time they spent in this mode, opting out of any serious discussions unless it was unavoidable.

"When you're in a setting where if you say something relevant, you can be perceived as smart or knowledgeable … It is to be educated, but also to kind of fit in with other people. In college, like every class I'm in, the first 5-10 minutes is talking about something relevant to today." (Prepping for Debate) – Neil (22, USA)

Finally, in Making Big Life Decisions, where participants sought to acquire knowledge to make important life decisions, they again applied the "like-minded, like-bodied" trust heuristic to home in on information that felt relevant to their lives, and trusted the wisdom of the crowd to guide their decision-making.

> "A few people said that the price for the program was not worth what you get out of it … So that's something that has stuck with me heavily." (Making Big Life Decision) — Claire (22, USA)

*4.1.1 Behavioral consequences of information modes*

Information modes reveal when and why different heuristics become relevant by highlighting how the trustworthiness of content matters in context. They reveal clear patterns in how young people are spending their time online, and why, which illuminates the limited extent to which traditional skills and heuristics for vetting and trusting information are activated.

Specifically, we found that participants spent the vast majority of their time in Timepass mode, where trustworthiness of content was less important than its entertainment value. It was a mode in which young people unconsciously considered trust and trust heuristics as irrelevant, and did not employ any literacy skills.

Literacy skills like vetting institutional credibility, expertise, and objectivity were relevant for only two of seven information modes, which were also two of the least popular modes: Keeping up with the Times and Prepping for Debate. Still, participants actively avoided these modes because they required engagement with news cycles, which participants found draining and upsetting; they had a high-mental tax (see Fig 2). As such, they applied heuristics that decreased the amount of time and energy they had to spend in these modes.

Maintaining a stable emotional state was an influential motivation for participants when navigating online spaces. We hypothesize that the desire to maintain a steady state explains why we observed participants spending more time in Timepass and Lifestyle Aspiration modes, but comparatively less in Prepping for Debate.

> "To be honest, if I'm not in the mood to see something serious, I'll keep scrolling … I don't want to start my day with negative energy." – Talia (24, USA)

Talia's phrasing leans on notions of infectious energies, moods, vibes and emphasizes the importance of maintaining mood over engaging with content that would disrupt it.

Participants would only engage in Prepping for Debate or Making a Big Life Decision modes for extremely important tasks. In these moments, they sought to complete their task as efficiently as possible so they could move back to Timepass where they felt emotionally soothed and stable.

We further suggest that this desire to maintain an emotional equilibrium was a response to information overload (Katz 2021). Our participants were exposed to a surfeit of digital content regarding complex global issues like climate change and war, which many found overwhelming. Strategic ignorance was a choice many of our participants made.

> If it's not something I know or at least have heard about already, it'll take a lot of effort to understand and won't be relevant to my life, [so I don't research it]. – Avery (20, USA)

Participants would switch quickly, almost automatically, between information modes to keep a steady state. Because participants were switching so fast between different modes, they were not always strict about their application of heuristics in one particular mode over another. They would often extend their attitude and behavior toward trust heuristics from Timepass into other modes.

> "I just like to read through [Facebook] ... Everybody my age will see [Facebook content] and be like, 'This is so ridiculous.' And not get angry with it ... I think a lot of us are just like, 'Oh my God, this is so funny that people will actually think that, it's so ridiculous.'"– Jamie (22, USA)

For Jamie, low quality content was a form of entertainment that she encountered with a relatively open and passive mindset. This left her vulnerable to inaccurate and harmful ideas. We observed these dynamics at play in many different cross-cultural scenarios. For example, Safia (24, IND) said she began watching Andrew Tate videos as entertainment and disagreed with his views. Over time, however, she adopted his views on traditional femininity and started to wear dresses more often.

While in Timepass mode, participants believed they simultaneously did not need heuristics for entertainment content and were impervious to misinformation because it was supposedly not to be taken seriously. When participants quickly shifted from Timepass to other modes, this opened them up to harm, as they would carry over a passive attitude of not needing to scrutinize information, inadvertently internalizing ideas. Especially when switching back and forth between Timepass and another mode, participants were thus not always consistently switching heuristics.

Our findings complicate the idea that users embark on linear information journeys armed with rationally activated media literacies. Rather than rationally completing linear information journeys, users are instead unconsciously and emotionally reacting as they wander through the variety of content available online.

*4.1.2 New trust heuristics of information modes*

Information modes revealed three new trust heuristics in participants' information seeking journeys:

1. **"Two-sideism":** Participants perceived content or creators who addressed both sides of an issue to be more trustworthy. When they could not find such sources and creators, they sought out two sources with extreme opposing views on an issue (e.g. pro-vaccination vs. anti-vaccination), and then decided what they believed. A few participants discussed two-sideism as a reaction against "media bias" or the polarization of society. We also see it as a reaction to post-truth society: the only way to arrive at the truth is to hear both sides of the story. For example, when in Prepping for Debate mode, many participants sought to learn two sides of an issue, especially when it involved politics. Since many doubted any source to be bias-free, they would parse two different sources that they believed covered opposing positions. This helped participants to feel that they were receiving unbiased information. For example, Tom (24, USA) would look for people online who think "the opposite way" (e.g. "anti-West radicals") to counteract what he believed to be mainstream pro-US views from The New York Times. He claimed to not trust these other views outright, but he liked having an "opposite perspective." By reviewing two opposing sides that he had chosen, Todd believed his final opinion to be more accurate.

2. **"Testing to trust":** Our participants were quick to act on claims they encountered not because they categorically believed them to be true, but because they wanted to test whether or not they were true. Instead of fact checking, they believed testing upon themselves to be the fastest and most accurate way to test for truth. This testing behavior was prominent in Lifestyle Aspiration and Getting Stuff Done modes. With eating and health content, participants would immediately try out diets and exercise routines they found online (Xu et al. 2024). As long as they were the only ones impacted and they believed themselves to be the best authority to judge truth (e.g., whether a diet or exercise routine is working for their body), they felt confident in their abilities to evaluate the content by acting as the "scientist" themselves.

3. **"Innate common sense":** Most participants felt their tech savviness, internalized knowledge, and quick processing abilities protected them from misinformation and from the potential harms of genAI. While heuristics are cues that enable quick evaluation of content, participants did not have awareness or concern about accuracy trade-offs they may be making in the process of using heuristics to assess veracity. This held true even when they acknowledged being fooled by misinformation previously. We observed participants judge video and text veracity after a few seconds, then retroactively explain their reaction in vague ways. For example, Neil (23, USA) originally judged an AI-generated photo of Donald Trump running away from police officers in New York as "real" because it was "common sense." This was because he already knew that Donald Trump was in New York City and was in criminal proceedings in the state. Neil fell for the genAI misinformation because it affirmed beliefs he had about the world.

## 4.2 GenAI as a new tool for info-seeking

With the introduction of genAI, we found that information modes shaped how the participants interacted with and trusted genAI tools.[5] For example, in Timepass mode, participants used genAI to generate fun, novel content to keep them entertained without any effort. In Lifestyle Aspiration Mode, they used genAI to generate practical plans and advice that they could quickly try out in their own lives.

| **Mode** | **Ways genAI is used in this mode** |
| --- | --- |
| Timepass | Generate fun, novel content |
| Lifestyle Aspiration | Provide advice/plans for users to try out to improve their lives |
| Bonding Over Content | Generate fun, novel, content to share with others |
| Getting Stuff Done | Do productivity tasks so users can quickly get back to "fun" things<br>*Note: "Fun" content as defined in Section 4.1 as "light" and often only involved the self* |
| Keeping Up With the Times | Summarize "important" issues<br>*Note: "Important" issues were defined as those where there was perceived social value to be aware and across the key details* |
| Prepping for Debate | Provide quick takes and links to other sources so users can have a "starting point" on "important" issues |
| Making a Big Life Decision | Get "unbiased" information on high-stakes decisions from finances to love<br>*Note: When making big life decisions, participants were particularly sensitive to ulterior motives, e.g. sponsored content* |

---

[5] At the time of this study, these were primarily ChatGPT 3.5 and Bard (now re-named Gemini).

Table 3. How we observed participants using genAI across the information mode

Notably, all participants were already adopting genAI across all information modes. At the time of the study, critics from popular press, civil society, government, and industry expressed concerns about the risks, dangers, and unreliability of genAI, and noted that young pepper were using genAI more than older generations (Ka Yuk Chan and Lee 2023; Lohr 2023; Metz 2023; Baxter and Schlesinger 2023; Sharma 2023). Despite intentionally recruiting for a sample of participants with varying degrees of familiarity with genAI, all of our participants were already using genAI or in the process of quickly adopting it. Participants saw genAI as an extension of their existing online practices – just another tool – and they were quick to use it. In fact, our participants felt an overwhelming pressure to become masters of genAI in order to remain competitive against their peers in school and on the job market.

In India, many of our participants were optimistic and excited about genAI—new technology was seen as key to upward social and economic mobility.

> "I do feel very amazed when I hear about all those [AI] things. Like, it's making your life very much easier … It'll increase your productivity … Thinking about the future, I feel very excited." – Gaurav (23, IND)

In the US, on the other hand, even as participants used genAI, many of them saw it as both a disruptor and competitor. Some feared that failing to master genAI would lead them to fall behind in school or be less competitive in their career field.

> "Who knows what the skill sets for future jobs will be, and if [knowing AI] will be a requirement. So it's kind of like, future proofing my capabilities … I just don't want to lose out on an opportunity."– Talia (24, USA)

Driven by this pressure to get ahead, participants typically valued applicability and efficiency over accuracy when evaluating genAI tools and its outputs. When prompted, participants expressed awareness that genAI results were not always accurate. However, they expressed comfort with "good enough" outputs, rationalizing that when it comes to genAI, perfect is the enemy of the good: they prized the tool's efficiency above all else.

> "When the quiz will be open for five minutes, we have ChatGPT in another window and quickly copy paste copy paste copy paste … Out of ten [questions], I got six correct. So, it's a pretty bad score … [But I still use ChatGPT], because at least you're getting six. In Google, you'll not get that. You'll get some random stuff, you'll not get answers." – Chana (20, IND)

Despite recognizing ChatGPT's low accuracy rate, Chana found its efficiency—specifically the way it formats information—to be directly applicable to her exams and too valuable to discontinue using the tool.

In India, many participants used genAI to surface direct answers, and were candid about copying genAI's answers directly. In the US, meanwhile, participants repeatedly emphasized that genAI was simply a "starting point" and insisted that they had editorial control over genAI. In reality, they would often adopt answers directly from genAI as their ownTalia (24, USA), for instance, had stated that she used Bard as only a "starting point" to find case studies for work. However, if she had never heard of the case study, rather than assume that Bard made it up, she doubted her own knowledge and instincts. She was predisposed to take Bard's answers at face-value, and did not exercise editorial control over the tool: she over-relied on genAI as a main source of information.

Participants were not oblivious to the risks and harms associated with genAI: most could articulate a few concerns. They simply perceived harms from technology as inevitable and outside of their ability to influence. As a result, rather than opting-out of technology, which they feared would negatively affect their future, or fight technology, which they perceived as futile, they preferred to master it. They acknowledged the harms of genAI, but the drive for efficiency overrode their concerns.

> "I don't think it's bad to use [genAI] for anything ... It's really useful and it's really time efficient. I feel like I don't lose trust [in genAI] because it's like, we know it's technology, and technology always has its downsides and limitations." – *Kathy, 21 (USA)*

In their rush to incorporate genAI quickly into their information seeking journeys, we observed three new trust heuristics specific to GenAI.

*4.2.1 New contexts, old heuristics*

Participants' desire to quickly adopt genAI to get ahead led them to import heuristics from other online domains that they felt confident with, namely, search engines and social media. In treating genAI as a new and improved version of these familiar domains, they imported the trust heuristics developed for these other technologies, ultimately forming flawed mental models of genAI and its capabilities.

Many participants used genAI as a new way to search for information online. In many ways, they felt it was a faster, more tailored search experience: they could ask a specific prompt and receive a single answer without having to sort through links. They often perceived the single answer as equally credible to search engine results, but faster and more accessible.

> "ChatGPT is pulling from this archive of information and sometimes that information could be stuff that's on Google. I used to rush to my phone to go to Safari, but now it's straight to ChatGPT." – Neil, 22 (USA)

In treating genAI like a search engine, they had formed a flawed mental model of how large language models (LLMs) worked. They imagined that the genAI tool was scanning the vast database of the internet and providing a synthesized top result. They assumed top answers could be afforded a high degree of trust and credibility because search results were perceived to filter and rank for quality.

Participants also imagined genAI as providing what we term a "social thermometer" of public opinion. Participants gauge a social thermometer by reading comments to understand how others are responding or if they are correcting what is being shared. This is a common practice on social media and a critical part of young people's opinion formation (Hassoun et. al. 2023). Since they misunderstood genAI to be scanning the internet, they imagined that it was distilling all thoughts shared online into a consensus opinion. As such, they perceived genAI as a "better" social thermometer relative to social media, synthesizing the wisdom of the crowd while saving them the energy of reading opinions themselves.

By importing trust heuristics from technologies where they felt in control, our participants quickly cemented their trust of genAI in familiar but flawed mental models, even though genAI outputs work differently and can be highly flawed.

*4.2.2 GenAI as subordinate to human control*

Participants felt strongly that genAI should be an unobtrusive tool that was subordinate to them. They wanted to be in control, and wanted genAI to respond in ways they could predict and understand. They implicitly wanted to assert that they still had mastery over functions that they saw as the domain of humans. They did not want genAI showing agency, authority, or providing unsolicited advice or information.

> "[ChatGPT] is informative, not a chatbot with its own thoughts and emotions. The more human mimicry that it attempts, the creepier it feels. None of us are using it to replace people." – Rhea (22, USA)

The desire for strict subordination of genAI was evident in many of the participants' early interactions with genAI. In initial interactions, they would test the genAI on a topic with which they were familiar and saw themselves as an authority on the subject. When trying out ChatGPT for the first time, Shanvi's (23, India) first query was about physiotherapy, which she knows a lot about as a physiotherapist. Testing genAI in this way helped participants assess the tool's accuracy, while they also felt secure knowing that the answer was not as detailed or thorough as their own knowledge. If the first test of knowledge was completed successfully, participants often rapidly concluded that genAI would most likely be correct or reliable on other topics, including those they were less familiar with. The fact that they were so quick to trust genAI that passed the test they set is another example of young people's trust in their testing methodology as described in Section 4.1.1.

Often, participants would follow up their initial query with a higher-level task to see genAI fail, thereby reinforcing participants' feelings of control and superiority over genAI. Neil (22, USA) purposely had ChatGPT take a college test that he was allowed to re-take, and took pride in outperforming ChatGPT:

> "I took an exam using answers only from AI and I got a 70%, which was horrible. But I had two attempts. So on my second attempt, I used my notes, I used my knowledge, and I got a 90%. So AI is advanced, but at least for this topic or this subject in school, it wasn't advanced enough." – Neil (22, USA)

Participants also sought to test genAI's boundaries with controversial prompts or inappropriate queries, such as asking for racist jokes. They did this not because they wanted the output; rather, they felt comfort in knowing what genAI's safeguards were – and that they were clever enough to outsmart these safeguards if they wanted. Testing the genAI product's boundaries was a way of establishing its limitations and confirming their human edge.

Through their tests of genAI, participants established both that it was accurate enough to incorporate into their lives *and* that it was non-threatening and "dumb." These qualities rendered genAI tools non-threatening, and made them feel in control of this powerful new technology.

*4.2.3 GenAI as a judgment-free zone*

Participants viewed genAI as a safe space where they could ask questions and conduct online activity without social risk or judgment. As is common for young people in this age cohort, they were keenly aware of and anxious about how they were perceived by their social circles. However, there may be a period effect whereby this age cohort now has heightened social pressures thanks to online interactions. Many participants expressed fear of social sanction or "being canceled," being embarrassed for not knowing the answer, or simply making social connections annoyed.

With genAI, young adults did not have to exercise their social communication skills to handle interpersonal conflict or differences of opinion. Neil (23, USA) was concerned about how studying abroad would impact his relationship with his girlfriend, but rather than speaking to her about it, he sought out discussion and tips for long-distance relationships from ChatGPT. Ismael (24, USA) used genAI to quickly learn personal finance terms that his wealthier peers used casually and he feared could judge him for not knowing, such as "liquidity pool" or "token sale." Talia (24, USA) could learn about fraught topics, such as Israel-Palestine and the Supreme Court decision on affirmative action, without having to know how to ask about questions on a sensitive topic. She used genAI tools like Bard to "educate" herself first before she discussed these topics with friends.

Participants in both India and US viewed genAI as judgment-free in part because it was perceived to be non-human. In India, our participants valued genAI for its confidentiality, unlike the potential risk of a human violating confidentiality. In the US, the primary draw of genAI was that it could not "cancel" them like their peers can. In both cases, using genAI to attenuate some of the potential social risk was appealing.

It is unsurprising that participants liked using genAI if it meant avoiding negative peer interactions, or feeling like they were a burden on their friends. Social mistakes were a source of the negative emotions they sought to avoid, and so, finding ways to avoid social mistakes or taking social risks altogether was a primary driver of their online behaviors. GenAI was another tool in service of this goal.

## 5 DISCUSSION

### 5.1 Implications of information modes for trust

We suggest that information modes have emerged as a coping mechanism in response to information overload. Many participants felt pressure to pay attention to their social media feeds and be on constant high alert, but they found all the information, especially news-related content, to be exhausting and emotionally overwhelming. It is a well-documented phenomenon that young adults are experiencing higher levels of uncertainty, anxiety, and stress compared to previous generations [Katz 2021]. While previous generations have experienced periods of global unrest, lost faith in institutions, and uncertainty, newer generations face periods of uncertainty alongside a barrage of real-time, on-the-ground footage and constant updates. Despite an increased awareness of mental health challenges, globally, individuals still bear more responsibility in protecting their own mental health than organizations, such as workplaces [Greenwood 2021; Breuning 2019; Sen 2020].

We posit that information modes are a response to both of these pressures: it is a "tactic" for managing information overload, and taking responsibility for one's own mental health [de Certeau 1984]. Switching between modes to achieve emotional equilibrium is an adaptive response young adults have developed to survive in this information landscape. In reaction to increasing global uncertainty and existential anxiety, our participants are spending most of their time in "light" modes, like Timepass and Lifestyle Aspiration.

This yields important implications for how social media and news platforms better safeguard their users from disinformation. The vast majority of information literacy interventions assume people are interested in the veracity of information, but we find that this is true only in very low-frequency modes. For example, information literacy interventions often encourage people to consider signals of accuracy or trustworthiness (such as evaluating the credibility of the source) or provide tips for literacy skills (such as lateral reading between sources) [Caufield 2017; Breakstone 2021]. However, we find that these tips and skills are relevant only when participants are either in the mindset of Prepping for Debate or Making a Big Life Decision—that is, when participants want to carefully consider different options. They minimized time in modes where media literacy skills are employed because they found these mentally taxing.

Due to the sheer amount of time spent in Timepass mode, when participants thoughtlessly and frequently switched to other modes, they would often apply their mindset and behaviors from Timepass to other modes they would switch to—before switching back to Timepass again. The overall effect is that participants lowered the bar to trust information even in contexts where they claim to care about veracity.

The paucity of time spent in modes concerned with veracity helps address a question that has long plagued information literacy research: why do information literacy interventions, such as fact-checking labels, often suffer from low usage [Clayton 2020]? Our findings shed some light on this phenomenon. Young people are not in the relevant information modes most of the time. Expecting people to engage with questions of trustworthiness and accuracy while they are in a completely different mindset (such as reminding people to fact check while in Timepass mode) means that at best, they ignore the intervention. At worst, they become irritated and disengage entirely.

Interventions must be aligned with a person's information mode to be effective. Participants perceived interventions that supported them in achieving the emotional goals of the mode they were in as helpful. For example, disclosures about an AI assistant's accuracy embedded in the output were perceived as helpful context in many modes, especially Prepping for Debate or Making a Big Life Decision. A warning banner on the other hand, was typically perceived as annoying as it was a barrier to achieving their goal.

### 5.2 Implications of genAI for trust

Participants' fast adoption of genAI across all information modes generated an array of blind spots and unintended consequences:

| Mode | Ways genAI is used in this mode | Unintended consequences (blind spot) |
| --- | --- | --- |
| Timepass | Generate fun, novel content | Convincing genAI content spreads quickly in mindless scrolling, including misinformation. |
| Lifestyle Aspiration | Provide advice/plans for users to try out to improve their lives | Participants were quick to test out AI-created action plans without first validating with other sources, essentially experimenting on their lives and their bodies with high consequences. |
| Bonding Over Content | Generate fun, novel, content to share with others | Convincing genAI content spreads quickly through value-based communities, including misinformation if it confirms existing beliefs. |
| Getting Stuff Done | Do boring tasks so users can quickly get back to "fun" things<br>*Note: "Fun" content as defined in Section 4.1 as "light" and often only involved the self* | Plagiarism of genAI content was normalized, whether explicitly or implicitly. |
| Keeping Up With the Times | Summarize "important" issues<br>*Note: "Important" issues were defined as those where there was perceived social value to be aware and across the key details* | GenAI summaries were treated as an authoritative source on complicated topics, because participants assumed it scanned the internet to generate content. |
| Prepping for Debate | Provide quick takes and links to other sources so users can have a "starting point" | GenAI would limit their purview to a more narrow "starting point" than they might have otherwise had from self-thinking, because genAI quick takes limited reflection time and encouraged "two-sideism" for every issue. |

| Making a Big Life Decision | Get "unbiased" information on decisions<br>*Note: When making big life decisions, participants were particularly sensitive to ulterior motives, e.g. sponsored content* | GenAI results reflected biases in how the prompt was phrased. |
|---|---|---|

Table 4. GenAI practices for each information mode and resulting unintended consequence (blindspot)

These blind spots resulted in three important implications and their potential negative social consequences:

*I. Trust in genAI outputs diminishes reflection before action*

The convenience and perceived accuracy in genAI outputs meant that participants spent less time searching and synthesizing, and acted upon information quickly. This can lead to two negative downstream effects. First, our participants' quick adoption of genAI answers as their own – whether it be to perform knowledge of a relevant topic, or to form their own opinion using genAI as a "starting point" – minimizes their exposure to heterogeneous sources, which may enhance groupthink and weaken critical thinking skills. Normalization of plagiarism further exacerbates this effect [Adhikari 2018; Handa 2005]. Moreover, in quickly moving to share or act on genAI answers, they also enable the spread of misinformation and its harms.

*II. Overconfidence leads to excessive "benefit of the doubt" for genAI*

In their initial interactions with genAI, participants worked to establish that they could "trick" it and perform higher-level tasks (like test-taking) better than genAI. This way, they rendered it non-threatening and established control over it. In seeking mastery of genAI—and in holding perceptions of genAI as non-threatening or as an "assistant" that performed low-level tasks—they often became overconfident, both in genAI and in themselves. Participants believed that genAI was scanning the internet to provide the single best answer, overestimating the accuracy of genAI results. They were also overly confident in their ability to instinctively know if something was incorrect, and that, if necessary, they would have the research skills to find the truth. The combination of both of these types of overconfidence meant that over time, they reflexively accepted what ChatGPT or Bard produced. In the event that participants did notice an error, they would often give genAI the benefit of the doubt, questioning their input rather than the model's output.

*III. Exacerbating some inequalities while closing others*

Some participants who previously felt excluded from elite knowledge used genAI as an equalizer. For example, Ismael (24, USA) described ChatGPT as a "teacher's aide" that has helped him learn at an "aggressive rate and in aggressive amounts" without feeling "embarrassed about [his] questions." In this way, genAI has helped him catch up to his more privileged peers. Other participants, however, saw genAI as another tool that helped the privileged get ahead. For example, Sydney (19, USA) told us that unlike her more privileged peers, she couldn't use ChatGPT to plagiarize because the risk of getting caught and losing her scholarship was too high. The effects of genAI on inequalities remains to be seen but will be dependent on local economic and social contexts as much as the models themselves.

## 6 CONCLUSION

Our study found that young people did not see genAI as distinct from other technologies, but rather, as a "super-charged" version of tools they were comfortable with like search engines and social media. To quickly master genAI, participants extended their current online info-seeking practices and heuristics in "information modes" to genAI products.

We suggest more research on the following topics to build on the findings of our study. Quantitative research of the ways young users spend time one (e.g. digital diary, device tracking) can validate the seven information modes we lay out in this paper and the relative amounts of time users spend in them, and help identify digital markers for users being in or switching to various information modes. These can be used to design interventions tailored to a specific mode. Empirical longitudinal research on the long-term impacts of individual interactions with genAI can provide more information about their social consequences, and reshape how we support digital literacy skills.

In order to design more effective interventions, we need to understand young people's information modes, and their desire for emotional equilibrium. This understanding will help us design better online spaces that make young people feel safe, affirmed, and in control, and encourage more positive relationships with genAI and the broader internet.

## Acknowledgements

We would like to thank Paree Zarolia, Ingrid Meintjes, Samar Elshafiey, Meena Natarajan, Jason Lipshin, and Amelia Hassoun for their comments and contributions to improve this research. We thank Michele Likely, Soojin Jeong, Yasmin Green, and Shira Almeleh for their reviews to improve this manuscript. Finally, we'd like to thank our participants for sharing their experiences with us.

# APPENDIX

## Appendix A. Demographic breakdowns of participants in both geographies

| New York Metro, USA | Participants | Bangalore, India | Participants |
|---|---|---|---|
| AI-expert | 5 | AI-expert | 4 |
| AI-familiar | 20 | AI-familiar | 22 |
| AI-unfamiliar | 1 | AI-unfamiliar | 2 |
| Rural | 4 | Rural | 3 |
| Urban or Suburban | 22 | Urban or Suburban | 22 |

| White | 11 | Hindu | 15 |
| Non-White (incl. black) | 16 | Muslim | 3 |
| Black | 4 | Other religion | 8 |
| Female | 14 | Female | 14 |
| Male | 10 | Male | 12 |
| Other gender | 2 | | |

**Appendix B. Summary Table of the Information Modes and the key motivations, emotions, actions, and heuristic**

| Mode | Motivation<br>*Why do they spend time online in this mode?* | Trust heuristic<br>*What key rules of thumb do they use while in this mode to quickly assess that content online is trustworthy?* | Representative quote<br>*What participant statement exemplifies how young people think while in this mode?* |
|---|---|---|---|
| Timepass | To assuage boredom and feel good while doing it | If it's entertainment, it doesn't matter if it's trustworthy or not | Gaurav (23, IND): "If it's just for humor, the fun entertainment industry and all, you can contradict yourself." |
| Lifestyle Aspiration | To imagine or work towards reaching ideals and improving yourself | • Creators who look and think like me are more trustworthy than experts<br>• It's faster and more reliable to test a lifestyle tip on myself than to research it<br>• GenAI can answer my doubts because it aggregates perspectives | Mehul (21, IND): "[Andrew Tate] might be wrong, he might be right, but the way he is influencing young men in society is quite motivating actually. [Because of him] I go to the gym more, work hard more…I want to get that rich." |
| Bonding over Content | To invest in online communities and unlock a sense of collective belonging | • Once I've joined a community that shares my values, I can automatically trust content that other members share<br>• Content shared in the community doesn't have to be rigorously fact-checked since it's just to bond over | Avery (20, USA): "There's a sense of community in [sharing info]. With fandom, it's not harmful to hype yourself up over something that doesn't end up being true…[But] sometimes the news sources are less informed than the fandom is. The fandom notices every little thing." |

| Getting stuff done | To complete annoying but necessary tasks with maximum efficiency | - GenAI is unbiased and accurate, because it scours the net and aggregates info<br>- Because I am just jump-starting my info journey, mostly for busywork, scrutiny can wait for later<br>- If I can imagine how to do this, or I can check what genAI is doing, I can just trust what genAI gives me | Irene (18, USA): "As long as I understand and I know what I'm talking about, at the end of the day it doesn't matter how I got there, especially if it's just like, stupid assignments." |
|---|---|---|---|
| Keeping up with the times | To stay up-to-date on important topics and feel like a good citizen | - Just knowing the bare minimum (e.g. headlines) is good enough to keep up<br>- It is easiest to agree with surrogate thinkers I've selected in the past<br>- I trust one-time-vetted sources to tell the truth If I see something several times, there must be truth to it | Paresh (21, IND): "I will wait for a trusted source to post [about a news event before reading it] …Each news source might have a different take, like, it's complicated. So I just stick to the ones I trust." |
| Prepping for debate | To learn just enough about topics that you feel others will judge you on later | - If it tells me 'both sides' of a topic, then I'm more inclined to trust<br>- The source has to be 'citable' for me to trust it, that is, others need to know and trust this source too<br>- I need to see the same info from multiple sources in different formats before I feel confident in my take<br>- Sponsorships or ads on serious content immediately means it can't be trusted | Neil, 22, USA: "When you're in a setting where if you say something relevant, you can be perceived as smart or knowledgeable …It is to be educated, but also to kind of fit in with other people. In college, like every class I'm in, the first 5-10 minutes is talking about something relevant to today." |
| Making big life decisions | To acquire additional knowledge to make an important life decision | - I trust an everyday person's perspective over "official" sources because they'll tell it to me straight<br>- I'm getting a visceral emotional reaction, so it must be true | Claire, 22, USA: "A few people said that the price for the program was not worth what you get out of it…So that's something that has stuck with me heavily." |

|  |  | - Unpolished visuals means emotions and reactions are more real, so I trust it more |  |